# An Effective Approach to Minimize Error in Midpoint Ellipse Drawing Algorithm

Dr. M. Javed Idrisi[1] and Aayesha Ashraf[2]

[1]Department of Mathematics
College of Natural and Computational Science
Mizan-Tepi University, Ethiopia
Email: mjavedidrisi@gmail.com

[2]Department of Computer Science and Engineering
Jamia Hamdard (Deemed to be University), New Delhi, India
Email: aayeshaa7@gmail.com

**Abstract**

The present paper deals with the generalization of Midpoint Ellipse Drawing Algorithm (MPEDA) to minimize the error in the existing MPEDA in cartesian form. In this method, we consider three different values of $h$, i.e., 1, 0.5 and 0.1. For $h = 1$, all the results of MPEDA have been verified. For other values of $h$ it is observed that as the value of $h$ decreases, the number of iteration increases but the error between the points generated and the original ellipse points decreases and vice-versa.

**Keywords:** Ellipse. Cartesian form of an Ellipse. MPEDA. Minimizing Error.

1. ## Introduction

A Midpoint Ellipse Drawing Algorithm (MPEDA) is used to determine the points needed for rasterizing an ellipse. In this algorithm, we divide the ellipse into 4 different quadrants and each quadrant will be divided into two regions $R_1$ and $R_2$, respectively. If we are able to plot the points in first quadrant, then by symmetry we can plot the points in the other three quadrants. Let $(x, y)$ be the point in first quadrant, then the points in other quadrants can be determined as shown in the given Table 1.

Table 1: Coordinates in the quadrants of an Ellipse

| Quadrant | I | II | III | IV |
|---|---|---|---|---|
| Point | $(x, y)$ | $(x, -y)$ | $(-x, -y)$ | $(-x, y)$ |

As we know that, every quadrant is again divided into two regions $R_1$ and $R_2$ (say). In order to complete a whole quadrant, we have to find out the points in the regions $R_1$ and $R_2$.

The equation of ellipse having center at origin is given by,

$$\frac{x^2}{a^2} + \frac{y^2}{b^2} = 1$$

where $a$ and $b$ are the semi major and minor axes respectively and $a > b$.

The above equation can be written as, $f(x,y) = 0$, where $f(x, y) = b^2x^2 + a^2y^2 - a^2b^2$. All points $(x, y)$ which satisfy the equation $f(x, y) = 0$ lies on the boundary of given ellipse. If $f(x, y) < 0$, the points will lie inside the ellipse and for $f(x, y) > 0$, the points will lie outside the ellipse, where $x$ and $y$ are real numbers.

**Initial Decision Parameter ($P'_k$) for the region $R_1$**

Let the initial point be $(0, b)$, i.e., $x_k = 0$ and $y_k = b$, therefore the initial decision parameter $P'_k$ is given by

$$P'_k = \frac{(4b^2 + a^2)}{4} - a^2b.$$

If $P'_k \geq 0$, then $y_{k+1} = y_k - 1$ and the next point will be $(x_k+1, y_k - 1)$, which gives

$$P'_{k+1} = P'_k + b^2 + 2b^2(x_k + 1) - 2a^2(y_k - 1)$$

If $P'_k < 0$, then $y_{k+1} = y_k$ and the next point will be $(x_k+1, y_k)$, which gives

$$P'_{k+1} = P'_k + b^2 + 2b^2(x_k + 1)$$

**Initial Decision Parameter ($P''_k$) for the region $R_2$**

Let $(x_k, y_k)$ be any point in the first region $R_2$, then the initial decision parameter $P''_k$ for the region $R_2$ is given by

$$P''_k = f(x_m, y_m) = b^2(x_k + 1/2)^2 + a^2(y_k - 1)^2 - a^2b^2.$$

If $P''_k \geq 0$, then $x_{k+1} = x_k$ and the next point will be $(x_k, y_k - 1)$, which gives

$$P''_{k+1} = P''_k + a^2 - 2a^2(y_k - 1)$$

If $P''_k < 0$, then $x_{k+1} = x_k + 1$ and the next point will be $(x_k+1, y_k - 1)$, which gives

$$P''_{k+1} = P''_k + b^2 + 2b^2(x_k + 1/2) + a^2 - 2a^2(y_k - 1)$$

The Ellipse drawing algorithm is studied by many researchers in past few decades. Some of the noteworthy work is as follows: Aken [1] has presented a midpoint algorithm for drawing ellipses on a raster graphics display which was highly accurate and only requires a few integer additions per pixel. They have also proved and demonstrated that a simple extension of J. E Bresenham's circle drawing algorithm doesn't guarantee accuracy in most general cases of ellipse. Hence the accuracy of the midpoint algorithm is limited only by the resolution of the display device itself, yet it requires no more execution time than the extended Bresenham algorithm. Kappel [2] has introduced an original algorithm for generating discrete approximations to ellipses for display on raster devices. This approach was not new, they have combined the existing ideas to render a better algorithm for evaluating from the benchmarks of efficiency, accuracy and elegance. Fellner et.al. [3] have studied an incremental and robust algorithm that efficiently computes the best approximation of general ellipses. Agathos et.al. [4] have described an algorithm similarly based on Bresenham methodology known as Efficient integer 8-connected algorithms for the fast generation of Conic Sections whose axes are aligned to the coordinate axes. Their performance results show that in the case of the ellipse, the algorithm is at least as fast as other known integer algorithms but requires lower integer range and always performs correct region transitions. In this algorithm antialiasing is easily incorporated. Liu et.al. [5] have developed a double-step circle drawing algorithm which gives the best approximate pixels to drawing a circle with only integer arithmetic. They showed that the speed of the algorithm introduced is higher than the existing circle drawing algorithms. Furthermore, their algorithm can be used to draw anti-aliased circles, e.g. to draw filled circle edges with different intensity, without increase in calculations. Haiwen et.al. [6] have proposed a hybrid algorithm for fast drawing ellipse. Dimri et.al. [7] have described midpoint ellipse algorithm as one of the popular algorithms for ellipse drawing. They have proposed an algorithm in that computes the pixels only in one octant and remaining part of the auxiliary circle can be generated by reflection about line $x = y$, with the help of the parametric equation of the ellipse.

The present work deals with the generalization of Midpoint Ellipse Drawing Algorithm (MPEDA) in Cartesian form. In this method, we consider three different values of $h$ i.e., 1, 0.5 and 0.1. For $h = 1$, all the results of MPEDA have been verified. For other values of $h$ it is observed that as the value of $h$ decreases, the number of iteration increases but the error between the points generated and the original ellipse points' decreases and vice-versa.

**2. Generalization of Midpoint Ellipse Drawing Algorithm**

The equation of ellipse having center at origin is given by,

$$\frac{x^2}{a^2} + \frac{y^2}{b^2} = 1 \qquad (1.1)$$

where *a* and *b* are the semi major and minor axes respectively and *a > b*.

The Eqn. (1.1) can be written as, $f(x,y) = 0$, where $f(x, y) = b^2x^2 + a^2y^2 - a^2b^2$. All points $(x, y)$ which satisfy the equation $f(x, y) = 0$ lies on the boundary of given ellipse. If $f(x, y) < 0$, the points will lie inside the ellipse and for $f(x, y) > 0$, the points will lie outside the ellipse, where $x$ and $y$ are real numbers.

**Initial Decision Parameter ($P'_k$) for the region $R_1$**

Let $(x_k, y_k)$ be any point in the first region $R_1$ and it is assumed that this point is moving in $xy$-plane in clockwise direction. Thus, the next point to $(x_k, y_k)$ in the region $R_1$ is given by $(x_k + h, y_k)$ or $(x_k + h, y_k - h)$, where $h$ is the width of the grid. Let $(x_m, y_m)$ be the mid-point of $(x_k + h, y_k)$ and $(x_k + h, y_k - h)$, therefore,

$$(x_m, y_m) = (x_k + h, y_k - h/2).$$

Let $P'_k$ be the value of the function $f(x, y)$ at the mid-point $(x_m, y_m)$, therefore,

$$P'_k = f(x_m, y_m) = b^2(x_k + h)^2 + a^2(y_k - h/2)^2 - a^2b^2. \tag{1.2}$$

Thus, $P'_{k+1}$ will be given by,

$$P'_{k+1} = b^2(x_{k+1} + h)^2 + a^2(y_{k+1} - h/2)^2 - a^2b^2. \tag{1.3}$$

where, $x_{k+1} = x_k + h$ and $y_{k+1} = y_k$ or $y_k - h$. Thus, form Eqn. (1.3), we have,

$$P'_{k+1} = b^2(x_k + 2h)^2 + a^2(y_{k+1} - h/2)^2 - a^2b^2. \tag{1.4}$$

On subtracting Eqn. (1.2) from Eqn. (1.4), we get,

$$P'_{k+1} - P'_k = \left[b^2(x_k + 2h)^2 + a^2(y_{k+1} - h/2)^2 - a^2b^2\right] - \left[b^2(x_k + h)^2 + a^2(y_k - h/2)^2 - a^2b^2\right].$$

On simplifying, we have,

$$P'_{k+1} - P'_k = b^2\left[(x_k + 2h)^2 - (x_k + h)^2\right] + a^2\left[(y_{k+1} - h/2)^2 - (y_k - h/2)^2\right]$$

which gives

$$P'_{k+1} = P'_k + b^2h^2 + 2b^2h(x_k + h) + a^2(y_{k+1}^2 - y_k^2) - a^2h(y_{k+1} - y_k) \tag{1.5}$$

Let the initial point be $(0, b)$, i.e., $x_k = 0$ and $y_k = b$, therefore from Eqn. (1.2), we get,

$$P'_k = \frac{(4b^2 + a^2)h^2}{4} - a^2bh. \tag{1.6}$$

If $P'_k \geq 0$, then $y_{k+1} = y_k - h$ and the next point will be $(x_k+h, y_k - h)$, which gives

$$P'_{k+1} = P'_k + b^2 h^2 + 2b^2 h(x_k + h) - 2a^2 h(y_k - h) \tag{1.7}$$

If $P'_k < 0$, then $y_{k+1} = y_k$ and the next point will be $(x_k+h, y_k)$, which gives

$$P'_{k+1} = P'_k + b^2 h^2 + 2b^2 h(x_k + h) \tag{1.8}$$

**Initial Decision Parameter $(P''_k)$ for the region $R_2$**

Let $(x_k, y_k)$ be any point in the first region $R_2$ then the next point to $(x_k, y_k)$ in the region $R_2$ is given by $(x_k, y_k - h)$ or $(x_k + h, y_k - h)$, where $h$ is the width of the grid. Let $(x_m, y_m)$ be the mid-point of $(x_k, y_k - h)$ or $(x_k + h, y_k - h)$, therefore,

$$(x_m, y_m) = (x_k + h/2, y_k - h).$$

Let $P''_k$ be the value of the function $f(x, y)$ at the mid-point $(x_m, y_m)$, therefore,

$$P''_k = f(x_m, y_m) = b^2(x_k + h/2)^2 + a^2(y_k - h)^2 - a^2 b^2. \tag{1.9}$$

Thus, $P''_{k+1}$ will be given by,

$$P''_{k+1} = b^2(x_{k+1} + h/2)^2 + a^2(y_{k+1} - h)^2 - a^2 b^2. \tag{1.10}$$

where, $x_{k+1} = x_k$ or $x_k + h$ and $y_{k+1} = y_k - h$. Thus, form Eqn. (1.10), we have,

$$P''_{k+1} = b^2(x_{k+1} + h/2)^2 + a^2(y_k - 2h)^2 - a^2 b^2. \tag{1.11}$$

On subtracting Eqn. (1.9) from Eqn. (1.11), we get,

$$P''_{k+1} - P''_k = b^2 \left[ (x_{k+1} + h/2)^2 - (x_k + h/2)^2 \right] + a^2 \left[ (y_k - 2h)^2 - (y_k - h)^2 \right]$$

On simplifying, we have,

$$P''_{k+1} = P''_k + b^2 \left[ (x_{k+1} + h/2)^2 - (x_k + h/2)^2 \right] + a^2 h^2 - 2a^2 h(y_k - h) \tag{1.12}$$

If $P''_k \geq 0$, then $x_{k+1} = x_k$ and the next point will be $(x_k, y_k - h)$, which gives

$$P''_{k+1} = P''_k + a^2 h^2 - 2a^2 h(y_k - h) \tag{1.13}$$

If $P''_k < 0$, then $x_{k+1} = x_k + h$ and the next point will be $(x_k+h, y_k - h)$, which gives

$$P^{"}_{k+1} = P^{"}_k + b^2h^2 + 2b^2h(x_k + h/2) + a^2h^2 - 2a^2h(y_k - h) \tag{1.14}$$

## 3. Results

Let us draw an ellipse having centre at (0, 0) and semi-major and minor axes are 8 units and 6 units, respectively by the general method of Midpoint ellipse drawing algorithm. We will plot the points only in the regions $R_1$ and $R_2$ of first quadrant considering (0, 6) as an initial point for different values of $h$.

**Table 2: Points in the region $R_1$ for $h = 1$**

| x | y | $P'_k$ | $x_{k+1}$ | $y_{k+1}$ | $P'_{k+1}$ | $2b^2x_{k+1}$ | $2a^2y_{k+1}$ |
|---|---|---|---|---|---|---|---|
| 0 | 6 | -332 | 1 | 6 | -224 | 72 | 768 |
| 1 | 6 | -224 | 2 | 6 | -44 | 144 | 768 |
| 2 | 6 | -44 | 3 | 6 | 208 | 216 | 768 |
| 3 | 6 | 208 | 4 | 5 | -108 | 288 | 640 |
| 4 | 5 | -108 | 5 | 5 | 288 | 360 | 640 |
| 5 | 5 | 288 | 6 | 4 | 244 | 432 | 512 |
| 6 | 4 | 244 | 7 | 3 | 400 | 504 | 384 |

In the last entry of Table 2, $2b^2x_{k+1} = 504 > 2a^2y_{k+1} = 384$. Thus, the region $R_2$ begins from this point and we will consider the point (7, 3) as an initial point for the region $R_2$ (Table 3). Similarly, for $h = 0.5$ the points are obtained in Table 4-5 and for $h = 0.1$ the points are given in Table 6-7.

**Table 3: Points in the region $R_2$ for $h = 1$**

| x | y | $P'_k$ | $x_{k+1}$ | $y_{k+1}$ | $P'_{k+1}$ | $2b^2x_{k+1}$ | $2a^2y_{k+1}$ |
|---|---|---|---|---|---|---|---|
| 7 | 3 | -23 | 8 | 2 | 361 | 576 | 256 |
| 8 | 2 | 361 | 8 | 1 | 297 | 576 | 128 |
| 8 | 1 | 297 | 8 | 0 | 361 | 576 | 0 |

**Table 4: Points in the region $R_1$ for $h = 0.5$**

| x | y | $P'_k$ | $x_{k+1}$ | $y_{k+1}$ | $P'_{k+1}$ | $2b^2x_{k+1}$ | $2a^2y_{k+1}$ |
|---|---|---|---|---|---|---|---|
| 0 | 6 | -179 | 0.5 | 6 | -152 | 36 | 768 |
| 0.5 | 6 | -152 | 1 | 6 | -107 | 72 | 768 |
| 1 | 6 | -107 | 1.5 | 6 | -44 | 108 | 768 |
| 1.5 | 6 | -44 | 2 | 6 | 37 | 144 | 768 |
| 2 | 6 | 37 | 2.5 | 5.5 | -216 | 180 | 704 |
| 2.5 | 5.5 | -216 | 3 | 5.5 | -99 | 216 | 704 |
| 3 | 5.5 | -99 | 3.5 | 5.5 | 36 | 252 | 704 |
| 3.5 | 5.5 | 36 | 4 | 5 | -131 | 288 | 640 |
| 4 | 5 | -131 | 4.5 | 5 | 40 | 324 | 640 |
| 4.5 | 5 | 40 | 5 | 4.5 | -59 | 360 | 576 |
| 5 | 4.5 | -59 | 5.5 | 4.5 | 148 | 396 | 576 |
| 5.5 | 4.5 | 148 | 6 | 4 | 117 | 432 | 512 |
| 6 | 4 | 117 | 6.5 | 3.5 | 136 | 468 | 448 |

### Table 5: Points in the region $R_2$ for $h = 0.5$

| $x$ | $y$ | $P'_k$ | $x_{k+1}$ | $y_{k+1}$ | $P'_{k+1}$ | $2b^2 x_{k+1}$ | $2a^2 y_{k+1}$ |
|---|---|---|---|---|---|---|---|
| 6.5 | 3 | -263.75 | 7 | 2.5 | -155.75 | 504 | 320 |
| 7 | 2.5 | -155.75 | 7.5 | 2 | 2.25 | 540 | 256 |
| 7.5 | 2 | 2.25 | 7.5 | 1.5 | -77.75 | 540 | 192 |
| 7.5 | 1.5 | -77.75 | 8 | 1 | 162.25 | 576 | 128 |
| 8 | 1 | 162.25 | 8 | 0.5 | 146.25 | 576 | 64 |
| 8 | 0.5 | 146.25 | 8 | 0 | 162.25 | 576 | 0 |

### Table 6: Points in the region $R_1$ for $h = 0.1$

| $x$ | $y$ | $P'_k$ | $x_{k+1}$ | $y_{k+1}$ | $P'_{k+1}$ | $2b^2 x_{k+1}$ | $2a^2 y_{k+1}$ |
|---|---|---|---|---|---|---|---|
| 0 | 6 | -37.88 | 0.1 | 6 | -36.8 | 7.2 | 768 |
| 0.1 | 6 | -36.8 | 0.2 | 6 | -35 | 14.4 | 768 |
| 0.2 | 6 | -35 | 0.3 | 6 | -32.48 | 21.6 | 768 |
| 0.3 | 6 | -32.48 | 0.4 | 6 | -29.24 | 28.8 | 768 |
| 0.4 | 6 | -29.24 | 0.5 | 6 | -25.28 | 36 | 768 |
| 0.5 | 6 | -25.28 | 0.6 | 6 | -20.6 | 43.2 | 768 |
| 0.6 | 6 | -20.6 | 0.7 | 6 | -15.2 | 50.4 | 768 |
| 0.7 | 6 | -15.2 | 0.8 | 6 | -9.08 | 57.6 | 768 |
| 0.8 | 6 | -9.08 | 0.9 | 6 | -2.24 | 64.8 | 768 |
| 0.9 | 6 | -2.24 | 1 | 6 | 5.32 | 72 | 768 |
| 1 | 6 | 5.32 | 1.1 | 5.9 | -61.92 | 79.2 | 755.2 |
| 1.1 | 5.9 | -61.92 | 1.2 | 5.9 | -52.92 | 86.4 | 755.2 |
| 1.2 | 5.9 | -52.92 | 1.3 | 5.9 | -43.2 | 93.6 | 755.2 |
| 1.3 | 5.9 | -43.2 | 1.4 | 5.9 | -32.76 | 100.8 | 755.2 |
| 1.4 | 5.9 | -32.76 | 1.5 | 5.9 | -21.6 | 108 | 755.2 |
| 1.5 | 5.9 | -21.6 | 1.6 | 5.9 | -9.72 | 115.2 | 755.2 |
| 1.6 | 5.9 | -9.72 | 1.7 | 5.9 | 2.88 | 122.4 | 755.2 |
| 1.7 | 5.9 | 2.88 | 1.8 | 5.8 | -58.04 | 129.6 | 742.4 |
| 1.8 | 5.8 | -58.04 | 1.9 | 5.8 | -44 | 136.8 | 742.4 |
| 1.9 | 5.8 | -44 | 2 | 5.8 | -29.24 | 144 | 742.4 |

| | | | | | | |
|---|---|---|---|---|---|---|
| 2 | 5.8 | -29.24 | 2.1 | 5.8 | -13.76 | 151.2 | 742.4 |
| 2.1 | 5.8 | -13.76 | 2.2 | 5.8 | 2.44 | 158.4 | 742.4 |
| 2.2 | 5.8 | 2.44 | 2.3 | 5.7 | -53.6 | 165.6 | 729.6 |
| 2.3 | 5.7 | -53.6 | 2.4 | 5.7 | -35.96 | 172.8 | 729.6 |
| 2.4 | 5.7 | -35.96 | 2.5 | 5.7 | -17.6 | 180 | 729.6 |
| 2.5 | 5.7 | -17.6 | 2.6 | 5.7 | 1.48 | 187.2 | 729.6 |
| 2.6 | 5.7 | 1.48 | 2.7 | 5.6 | -50.4 | 194.4 | 716.8 |
| 2.7 | 5.6 | -50.4 | 2.8 | 5.6 | -29.88 | 201.6 | 716.8 |
| 2.8 | 5.6 | -29.88 | 2.9 | 5.6 | -8.64 | 208.8 | 716.8 |
| 2.9 | 5.6 | -8.64 | 3 | 5.6 | 13.32 | 216 | 716.8 |
| 3 | 5.6 | 13.32 | 3.1 | 5.5 | -34.4 | 223.2 | 704 |
| 3.1 | 5.5 | -34.4 | 3.2 | 5.5 | -11 | 230.4 | 704 |
| 3.2 | 5.5 | -11 | 3.3 | 5.5 | 13.12 | 237.6 | 704 |
| 3.3 | 5.5 | 13.12 | 3.4 | 5.4 | -31.16 | 244.8 | 691.2 |
| 3.4 | 5.4 | -31.16 | 3.5 | 5.4 | -5.6 | 252 | 691.2 |
| 3.5 | 5.4 | -5.6 | 3.6 | 5.4 | 20.68 | 259.2 | 691.2 |
| 3.6 | 5.4 | 20.68 | 3.7 | 5.3 | -20.16 | 266.4 | 678.4 |
| 3.7 | 5.3 | -20.16 | 3.8 | 5.3 | 7.56 | 273.6 | 678.4 |
| 3.8 | 5.3 | 7.56 | 3.9 | 5.2 | -30.56 | 280.8 | 665.6 |
| 3.9 | 5.2 | -30.56 | 4 | 5.2 | -1.4 | 288 | 665.6 |
| 4 | 5.2 | -1.4 | 4.1 | 5.2 | 28.48 | 295.2 | 665.6 |
| 4.1 | 5.2 | 28.48 | 4.2 | 5.1 | -6.2 | 302.4 | 652.8 |
| 4.2 | 5.1 | -6.2 | 4.3 | 5.1 | 25.12 | 309.6 | 652.8 |
| 4.3 | 5.1 | 25.12 | 4.4 | 5 | -6.84 | 316.8 | 640 |
| 4.4 | 5 | -6.84 | 4.5 | 5 | 25.92 | 324 | 640 |
| 4.5 | 5 | 25.92 | 4.6 | 4.9 | -3.32 | 331.2 | 627.2 |
| 4.6 | 4.9 | -3.32 | 4.7 | 4.9 | 30.88 | 338.4 | 627.2 |
| 4.7 | 4.9 | 30.88 | 4.8 | 4.8 | 4.36 | 345.6 | 614.4 |
| 4.8 | 4.8 | 4.36 | 4.9 | 4.7 | -20.16 | 352.8 | 601.6 |
| 4.9 | 4.7 | -20.16 | 5 | 4.7 | 16.2 | 360 | 601.6 |
| 5 | 4.7 | 16.2 | 5.1 | 4.6 | -5.6 | 367.2 | 588.8 |
| 5.1 | 4.6 | -5.6 | 5.2 | 4.6 | 32.2 | 374.4 | 588.8 |
| 5.2 | 4.6 | 32.2 | 5.3 | 4.5 | 13.12 | 381.6 | 576 |
| 5.3 | 4.5 | 13.12 | 5.4 | 4.4 | -3.96 | 388.8 | 563.2 |
| 5.4 | 4.4 | -3.96 | 5.5 | 4.4 | 36 | 396 | 563.2 |
| 5.5 | 4.4 | 36 | 5.6 | 4.3 | 21.64 | 403.2 | 550.4 |
| 5.6 | 4.3 | 21.64 | 5.7 | 4.2 | 9.28 | 410.4 | 537.6 |
| 5.7 | 4.2 | 9.28 | 5.8 | 4.1 | -1.08 | 417.6 | 524.8 |
| 5.8 | 4.1 | -1.08 | 5.9 | 4.1 | 41.76 | 424.8 | 524.8 |
| 5.9 | 4.1 | 41.76 | 6 | 4 | 34.12 | 432 | 512 |
| 6 | 4 | 34.12 | 6.1 | 3.9 | 28.48 | 439.2 | 499.2 |
| 6.1 | 3.9 | 28.48 | 6.2 | 3.8 | 24.84 | 446.4 | 486.4 |
| 6.2 | 3.8 | 24.84 | 6.3 | 3.7 | 23.2 | 453.6 | 473.6 |

| x | y | $P'_k$ | $x_{k+1}$ | $y_{k+1}$ | $P'_{k+1}$ | $2b^2 x_{k+1}$ | $2a^2 y_{k+1}$ |
|---|---|---|---|---|---|---|---|
| 6.3 | 3.7 | 23.2 | 6.4 | 3.6 | 23.56 | 460.8 | 460.8 |
| 6.4 | 3.6 | 23.56 | 6.5 | 3.5 | 25.92 | 468 | 448 |

**Table 7: Points in the region $R_2$ for $h = 0.1$**

| x | y | $P'_k$ | $x_{k+1}$ | $y_{k+1}$ | $P'_{k+1}$ | $2b^2 x_{k+1}$ | $2a^2 y_{k+1}$ |
|---|---|---|---|---|---|---|---|
| 6.5 | 3.5 | -19.67 | 6.6 | 3.4 | -15.03 | 475.2 | 435.2 |
| 6.6 | 3.4 | -15.03 | 6.7 | 3.3 | -8.39 | 482.4 | 422.4 |
| 6.7 | 3.3 | -8.39 | 6.8 | 3.2 | 0.25 | 489.6 | 409.6 |
| 6.8 | 3.2 | 0.25 | 6.8 | 3.1 | -38.79 | 489.6 | 396.8 |
| 6.8 | 3.1 | -38.79 | 6.9 | 3 | -26.87 | 496.8 | 384 |
| 6.9 | 3 | -26.87 | 7 | 2.9 | -12.95 | 504 | 371.2 |
| 7 | 2.9 | -12.95 | 7.1 | 2.8 | 2.97 | 511.2 | 358.4 |
| 7.1 | 2.8 | 2.97 | 7.1 | 2.7 | -30.95 | 511.2 | 345.6 |
| 7.1 | 2.7 | -30.95 | 7.2 | 2.6 | -11.75 | 518.4 | 332.8 |
| 7.2 | 2.6 | -11.75 | 7.3 | 2.5 | 9.45 | 525.6 | 320 |
| 7.3 | 2.5 | 9.45 | 7.3 | 2.4 | -20.63 | 525.6 | 307.2 |
| 7.3 | 2.4 | -20.63 | 7.4 | 2.3 | 3.85 | 532.8 | 294.4 |
| 7.4 | 2.3 | 3.85 | 7.4 | 2.2 | -23.67 | 532.8 | 281.6 |
| 7.4 | 2.2 | -23.67 | 7.5 | 2.1 | 4.09 | 540 | 268.8 |
| 7.5 | 2.1 | 4.09 | 7.5 | 2 | -20.87 | 540 | 256 |
| 7.5 | 2 | -20.87 | 7.6 | 1.9 | 10.17 | 547.2 | 243.2 |
| 7.6 | 1.9 | 10.17 | 7.6 | 1.8 | -12.23 | 547.2 | 230.4 |
| 7.6 | 1.8 | -12.23 | 7.7 | 1.7 | 22.09 | 554.4 | 217.6 |
| 7.7 | 1.7 | 22.09 | 7.7 | 1.6 | 2.25 | 554.4 | 204.8 |
| 7.7 | 1.6 | 2.25 | 7.7 | 1.5 | -16.31 | 554.4 | 192 |
| 7.7 | 1.5 | -16.31 | 7.8 | 1.4 | 22.57 | 561.6 | 179.2 |
| 7.8 | 1.4 | 22.57 | 7.8 | 1.3 | 6.57 | 561.6 | 166.4 |
| 7.8 | 1.3 | 6.57 | 7.8 | 1.2 | -8.15 | 561.6 | 153.6 |
| 7.8 | 1.2 | -8.15 | 7.9 | 1.1 | 35.29 | 568.8 | 140.8 |
| 7.9 | 1.1 | 35.29 | 7.9 | 1 | 23.13 | 568.8 | 128 |
| 7.9 | 1 | 23.13 | 7.9 | 0.9 | 12.25 | 568.8 | 115.2 |
| 7.9 | 0.9 | 12.25 | 7.9 | 0.8 | 2.65 | 568.8 | 102.4 |
| 7.9 | 0.8 | 2.65 | 7.9 | 0.7 | -5.67 | 568.8 | 89.6 |
| 7.9 | 0.7 | -5.67 | 8 | 0.6 | 44.89 | 576 | 76.8 |
| 8 | 0.6 | 44.89 | 8 | 0.5 | 39.13 | 576 | 64 |
| 8 | 0.5 | 39.13 | 8 | 0.4 | 34.65 | 576 | 51.2 |
| 8 | 0.4 | 34.65 | 8 | 0.3 | 31.45 | 576 | 38.4 |
| 8 | 0.3 | 31.45 | 8 | 0.2 | 29.53 | 576 | 25.6 |
| 8 | 0.2 | 29.53 | 8 | 0.1 | 28.89 | 576 | 12.8 |

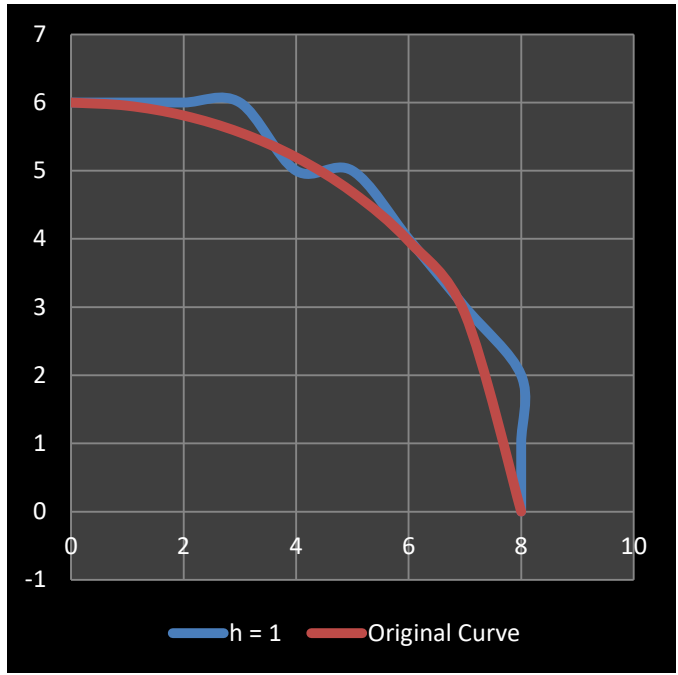

**Fig. 1: Original curve versus curve for *h* = 1 in the first quadrant**

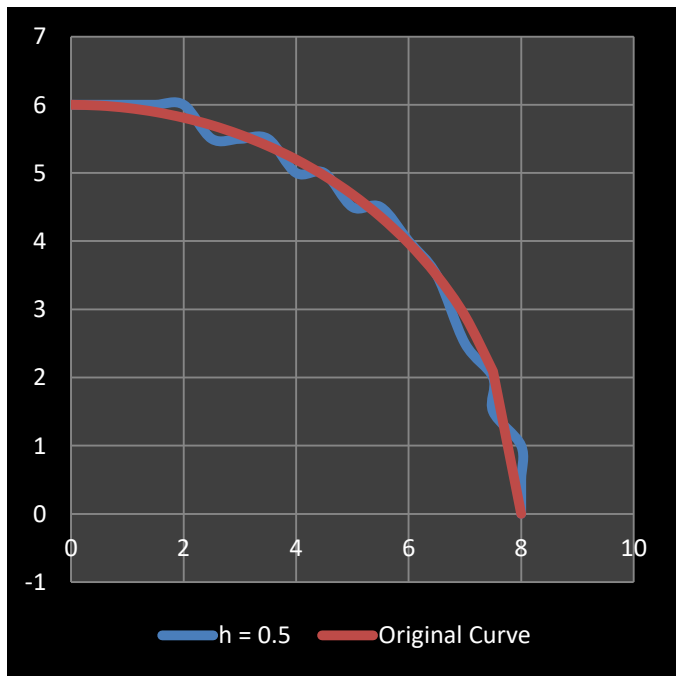

**Fig. 2: Original curve versus curve for *h* = 0.5 in the first quadrant**

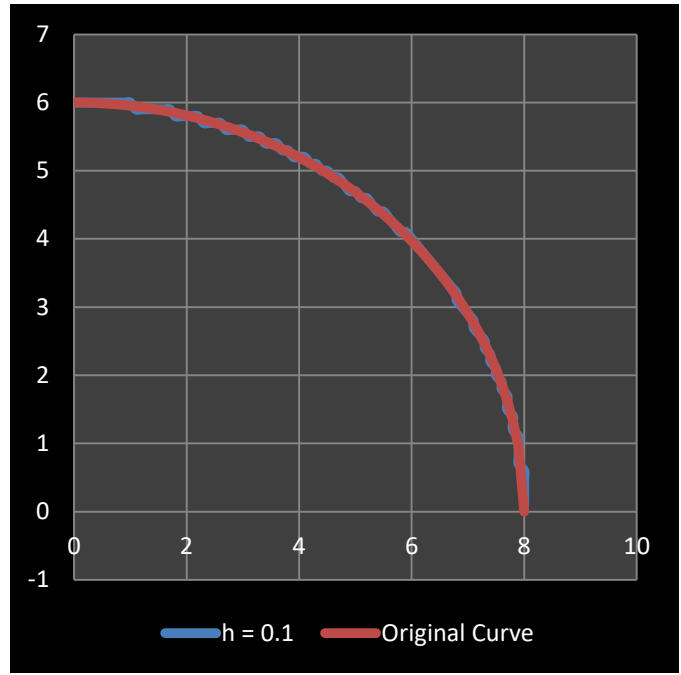

**Fig. 3: Original curve versus curve for *h* = 0.1 in the first quadrant**

4. Discussion

The number of iterations depends upon the value of *h*. As *h* decreases, the number of iteration increases showed in Table 2, 3, 4, 5, 6 and 7. We have solved the same problem considering three different values of *h*. For *h* = 1, the classical case of MPEDA has been verified and it is clearly shown in Fig. 1 that there is a wide difference between the original curve and the curve generated by MPEDA. In the next Table, i.e., Table 4, we considered the value of *h* as 0.5 though the number of iterations increased but the results are much closer to original one (Fig. 2). Now, we considered *h* as 0.1, we observed that the points are almost similar to the original curve (Fig. 3).

**Declaration of Competing Interest**

The authors declare that there is no conflict of interests regarding the publication of this manuscript.